\documentclass[sigconf, nonacm=true]{acmart}

\AtBeginDocument{%
  }

\copyrightyear{2023}
\acmYear{2023}
\setcopyright{acmlicensed}\acmConference[IH\&MMSec '23]{Proceedings of the
	2023 ACM Workshop on Information Hiding and Multimedia Security}{June
	28--30, 2023}{Chicago, IL, USA}
\acmBooktitle{Proceedings of the 2023 ACM Workshop on Information Hiding
	and Multimedia Security (IH\&MMSec '23), June 28--30, 2023, Chicago, IL, USA}
\acmPrice{15.00}
\acmDOI{10.1145/3577163.3595093}
\acmISBN{979-8-4007-0054-5/23/06}

\usepackage{tikz}
\usepackage{xcolor}
\usetikzlibrary{shapes.geometric, arrows}

\tikzstyle{box} = [rectangle, text centered]
\tikzstyle{arrow} = [thick,->,>=stealth]
\tikzstyle{textbox} = [rectangle, text centered, draw=black, fill={rgb, 255:red, 249; green, 231; blue, 202}, draw={rgb, 255:red, 242; green, 202; blue, 140}, line width=1.5]

\begin{document}

\title{Compatibility and Timing Attacks for JPEG Steganalysis}

\author{Etienne Levecque}

\affiliation{%
  \institution{Univ. Lille, CNRS, Centrale Lille,
  	UMR 9189 CRIStAL}
  \city{Lille}
  \country{France}
}
\email{etienne.levecque@cnrs.fr}

\author{Patrick Bas}
\affiliation{%
	\institution{Univ. Lille, CNRS, Centrale Lille,
		UMR 9189 CRIStAL}
	\city{Lille}
	\country{France}
}
\email{patrick.bas@cnrs.fr}

\author{Jan Butora}

\affiliation{%
	\institution{Univ. Lille, CNRS, Centrale Lille,
		UMR 9189 CRIStAL}
	\city{Lille}
	\country{France}
}
\email{jan.butora@cnrs.fr}

\renewcommand{\shortauthors}{Etienne Levecque, Patrick Bas, Jan Butora}

\begin{abstract}
	This paper introduces a novel compatibility attack to detect a steganographic message embedded in the DCT domain of a JPEG image at high-quality factors (close to 100).
	Because the JPEG compression is not a surjective function, {\it i.e.} not every DCT blocks can be mapped from a pixel block, embedding a message in the DCT domain can create incompatible blocks. We propose a method to find such a block, which directly proves that a block has been modified during the embedding.
	This theoretical method provides many advantages such as being completely independent to Cover Source Mismatch, having good detection power, and perfect reliability since false alarms are impossible as soon as incompatible blocks are found.
	We show that finding an incompatible block is equivalent to proving the infeasibility of an Integer Linear Programming problem.
	However, solving such a problem requires considerable computational power and has not been reached for 8x8 blocks.
	Instead, a timing attack approach is presented to perform steganalysis without potentially any false alarms for large computing power.
\end{abstract}


\begin{teaserfigure}
  \includegraphics[width=\textwidth]{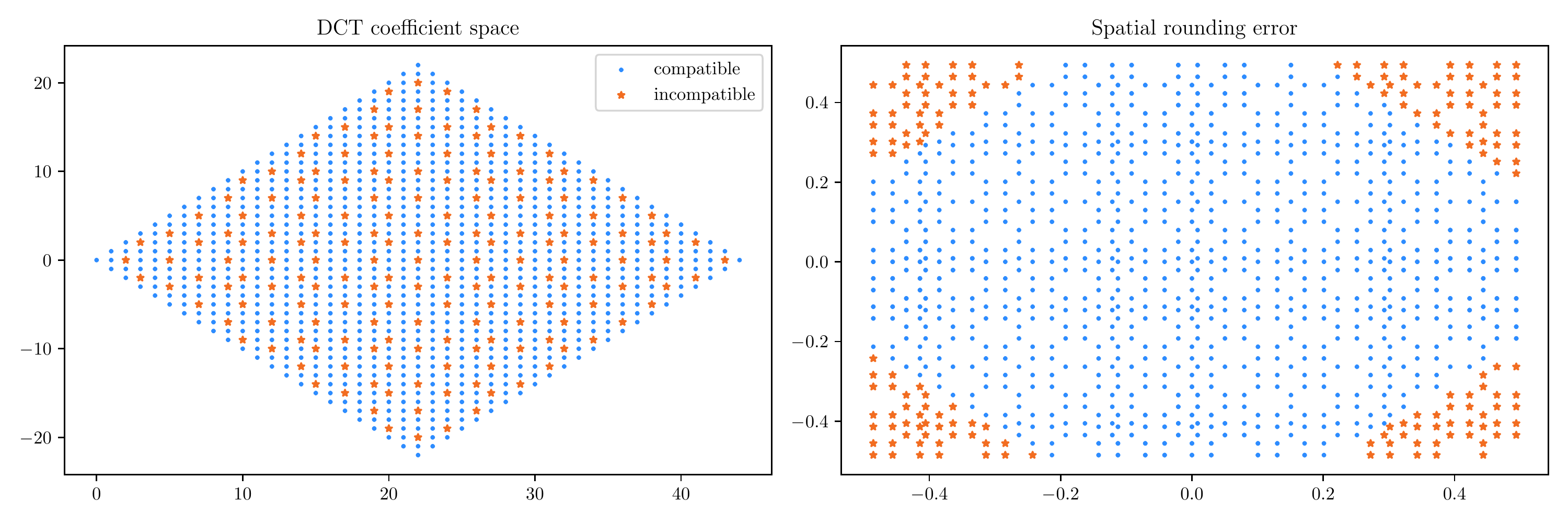}
  \caption{Toy example using blocks of size $(1,2)$ for visualization purposes. Blue dots are all the compatible DCT blocks obtained by compressing pairs of values between 0 and 32. The orange stars are the DCT blocks without any antecedent in the pixel domain, thus incompatible and stego. The distribution of the spatial rounding errors (left plot) can be used to detect incompatible blocks, however for larger block sizes, regions of incompatible errors are more complex.
  }
  \label{fig:banner}
\end{teaserfigure}


\maketitle

\section{Introduction and related works}

Steganography is the art of concealing a secret message inside innocuous media to hide communication from everyone but the recipient. The media used to hide the exchange is called a cover and most digital media such as images and video files are well suited for this task. Indeed, they can carry lots of bytes of information that are irrelevant to humans and often difficult to detect for a machine.

Compatibility attacks are one type of algorithm in the toolbox of steganalysis experts. They are considered universal in the sense that they do not target a special embedding algorithm, but a format of cover media. The main idea is to find a property present in every cover of the same format, if this property is not present in a file, then we can deduce, with some degree of incertitude, that this file hides a message. 
\subsection{Related works}

The JPEG format uses the Discrete Cosine Transform (DCT) and rounding functions to compress an image. Those operations create useful properties for compatibility attacks. The literature about such attacks on JPEG images can be divided into two main categories: one based on hard properties and one based on statistical properties. 

One example of hard property can be found in one of the first papers on digital image steganalysis by Fridrich {\it et al.}~\cite{fridrich2001steganalysis}, which focuses on messages embedded in the pixel domain for JPEG decompressed images. Such an embedding scheme can create incompatible blocks because they do not have any mathematical inverse in the DCT domain. A brute force approach is proposed to detect such blocks but is limited to quality factors lower than 95 because of complexity limitation. As a side note, the approach we propose in this paper is conceptually similar to this pioneering reference, but contrary to the reference, the embedding is also performed in the JPEG domain and the property is also different. A similar strategy relying on the detection of incompatible JPEG blocks was proposed later on by Yousfi {\it et al.}~\cite{yousfi2019breaking}. In this paper, authors compute the feasible dynamic range of DCT DC coefficients to detect a few stego images that have been embedded in the DCT domain because the DC value is larger than $1016$. Note that the main advantage of compatibility attacks using hard properties is perfect reliability, in the sense that a false alarm (i.e. detecting a stego content when there is none) is impossible.

Examples of statistical properties are given by Kodovsky {\it et. al.}~\cite{kodovsky2013jpeg}, who target messages embedded in the pixel domain for every quality factor. A feature is extracted and then classified using a fusion of Fisher Discriminant Analysis. A more recent example has been proposed by Butora {\it et. al.}~\cite{butora2019reverse} where authors target statistical properties present in a feature called spatial rounding error. The method focuses on messages embedded in the DCT coefficients of JPEG images compressed at a high-quality factor (99 or 100). This feature has a different distribution depending on the presence of a message or not. The variance of the rounding error or CNN-based steganalysis are then used to distinguish between both classes.  

\subsection{Outline of the paper}

The compatibility attack proposed in this paper can detect a message embedded in the DCT coefficients for high-quality factor compression (close to 100) and belongs to the first category: it focuses on a hard property. 

The property used is the non-surjectivity of the compression: every DCT block in a cover image must have an inverse image in the pixel domain. But if this block is modified, the latter assertion can sometimes no longer be true. We call them "incompatible" blocks since no integer antecedent exists in the spatial domain. A toy visualization is shown in figure \ref{fig:banner} where those incompatible blocks are orange stars and compatible DCT blocks are blue dots. Remark how the spatial rounding error is a very good feature because it perfectly separates the space between compatible and incompatible blocks. The presence of incompatible blocks could explain the excellent detection results obtained by Butora {\it et. al.}~\cite{butora2019reverse} who are training a Deep Learning model on potentially incompatible rounding errors. Our method do not use Deep Learning but profits from perfect reliability and also benefits from good detection power because there are potentially a lot of incompatible blocks in a stego image, and a single incompatible block in the images is enough to classify it as stego.

To show that a block is incompatible, we show that the inverse image of a DCT block is a solution to an Integer Linear Programming (ILP) problem. If this problem is proven to be unfeasible it means that the block does not have any inverse image, hence being incompatible. However, proving that such a problem is infeasible for $8\times8$ blocks has shown complexity difficulties. For this reason and clarity purposes, a sandbox approach of the JPEG with smaller block sizes will be used to explain the concept and build the detector. Moreover, even if the incompatibility cannot be practically shown for $8\times8$ stego blocks with our current computational power, finding a solution for cover blocks takes a reasonable amount of time. Therefore it is possible to derive a timing attack to perform steganalysis on real JPEG images. 

In cryptography, this methodology, called "Timing attacks", is a side-channel attack used to infer the key (see for example~\cite{kocher1996timing}). It is based on the fact that the time necessary to encrypt a message can be dependent on the composition of the key. In this paper, the timing attack is intertwined with the compatibility attack since we will show in the rest of the paper that the time to find compatible blocks can also be used as a steganalysis feature.

The first section will introduce JPEG compression and its corresponding notations. The second section will focus on the characterization of the non-surjectivity property as well as its formulation as an Integer Linear Programming problem. The third section will present two sandbox experiments to analyze those incompatible blocks for different sizes of JPEG blocks. Finally, the last section will present a steganalysis detector based on a timing attack for $8\times8$ blocks.

\section{Preliminaries}

\subsection{Notations}

Bold symbols will be used to identify vectors and matrices only. The notation $[\cdot]$ is the rounding function towards an integer. JPEG compression uses 2D blocks, however for clarity purpose, we prefer to work with flattened version of the blocks, thus identifying the coordinate of a block with a single index: $\mathbf{x} = (x_i)$. Concerning indexes, $i$ is reserved to coordinate in the spatial domain (also called pixel domain) and $k$ is reserved for the DCT domain (also called frequency domain).

\subsection{JPEG Compression}\label{sec:jpeg}

We start by introducing a generalized version of the JPEG compression for an arbitrary 2D block size $(n,m)$ at a quality factor of 100. At such a quality factor, the quantization table is equal to 1 for every coefficient, hence we will ignore it in the following notations. Let $\mathbf{M} \in \mathcal{M}_{nm,nm}(\mathbb{R})$ be the matrix associated with the Discrete Cosine Transform, a linear transformation that can be considered as a rotation. The inverse DCT is simply the transpose of the matrix: $\mathbf{M}^T$. The following notations will be the same throughout the paper:
\begin{alignat*}{3}
	\mathbf{x} & &\in \left[0;255\right]^{nm} \qquad &\text{Original integer block flattened,}\\
	\mathbf{d} &= \mathbf{M} \mathbf{x} \quad &\in \mathbb{R}^{nm} \qquad &\text{Floating DCT coefficients,}\\
	\mathbf{c} &= [\mathbf{d}] &\in \mathbb{Z}^{nm} \qquad &\text{Integer DCT coefficients,}\\
	\mathbf{y} & = \mathbf{M}^T \mathbf{c} & \in \mathbb{R}^{nm} \qquad &\text{Floating decompressed block.}
\end{alignat*}
The final decompressed block can be obtained by rounding and clipping $\mathbf{y}$.

The JPEG pipeline is lossy during compression even for a quality factor of 100. This is due to the rounding function applied in the DCT and spatial domain. Let us define those two errors:
\begin{align*}
\mathbf{e} & = [\mathbf{y}] - \mathbf{y} \qquad \text{Rounding error in spatial domain,}\\
\mathbf{u} & = [\mathbf{d}] - \mathbf{d} \qquad \text{Rounding error in DCT domain.}
\end{align*}

JPEG compression does not always use the same rounding function and the function used is not explicitly stored in the image. However, some methods such as the one presented in Butora {\it et. al.}~\cite{butora2022compressor} allow us to detect the rounding function used at high-quality factor. In our case, we use the round toward the nearest integer and especially the version round half away from zero. This function is the one used in the very popular JPEG library \texttt{libjpeg} as explained by Bene\v{s} {\it et. al.}~\cite{benes2022libjpeg}. 

With those notations, we will see in the next section that JPEG compression is not a surjective function and we will characterize compatible DCT blocks using the spatial rounding error.

\section{Non-surjectivity of JPEG compression}

\subsection{Definition of the non-surjectivity:}

It is defined by the fact that there exist DCT blocks $\mathbf{c} \in \mathbb{Z}^{nm}$, such that there is no pixel block $\mathbf{\tilde{x}} \in \left[0;255\right]^{nm}$ for which $c=\left[\mathbf{M\tilde{x}}\right]$.

In other words, such blocks $\mathbf{c}$ do not have any antecedent\footnote{also called "inverse image" or "preimage" in algebra.} and those blocks are considered incompatible. Note that if a block is compatible it can have multiple antecedents, that is why we use the notation $\tilde{\cdot}$ to identify a compatible solution.

Given a DCT block $\mathbf{c}$, if we want to find one compatible antecedent, the known variables are $\mathbf{e}$ and $\mathbf{y}$. The main unknown variable is the DCT rounding error $\mathbf{u}$. Thus, we start by deriving a characterization of such rounding error:
\begin{equation}\label{eq1}
	\begin{split}
		[\mathbf{y}] - \mathbf{x}& = [\mathbf{y}] - \mathbf{M}^T\mathbf{d},\\
		& = [\mathbf{y}] - \mathbf{M}^T(\mathbf{c} - \mathbf{u}),\\
		& = [\mathbf{y}] - \mathbf{M}^T\mathbf{c} + \mathbf{M}^T\mathbf{u},\\
		& = [\mathbf{y}] - \mathbf{y} + \mathbf{M}^T\mathbf{u}, \\
		& = \mathbf{e} + \mathbf{M}^T\mathbf{u}.
	\end{split}
\end{equation}

Finally, by defining $\mathbf{k} = \mathbf{e} + \mathbf{M}^T\mathbf{u}$ we have that $\mathbf{u} = \mathbf{M}(\mathbf{k} - \mathbf{e})$. Note that the first part of this equation is a vector of integer, thus forcing $\mathbf{k}$ to also be an integer. Without loss of generality, the following characterization is defined for the rounding half-down function.

\subsection{Characterization:}

Let $\mathbf{k} \in \mathbb{Z}^{nm}$. We define $\mathbf{\tilde{x}} = [\mathbf{y}] - \mathbf{k}$ and $\mathbf{\tilde{u}} = \mathbf{M}(\mathbf{k} - \mathbf{e})$ the DCT rounding error associated to the compression of $\mathbf{\tilde{x}}$. $\mathbf{\tilde{x}}$ is a compatible antecedent of $\mathbf{c}$ if and only if

\begin{equation}\label{constraints}
\left\{\begin{split}
-0.5 < \tilde{u}_i \leq 0.5, &\quad \text{ if } d_i \geq 0,\\
-0.5 \leq \tilde{u}_i < 0.5, &\quad \text{ if } d_i < 0.
\end{split}\right.
\end{equation}

Now that we have a condition to identify compatible antecedents, we would like to find them. However, the set of possibilities depends on every value of $\mathbf{k}$. We can observe empirically (using the compression error of real images from the ALASKA2 dataset) that for quality factor 100, $\mathbf{k} \in \left\{-1,0,1\right\}^{nm}$ which gives us an upper bound of the possibility of $3^{nm}$. Hence, the brute force approach is impossible for $8 \times 8$ blocks. 

\subsection{Optimization problem}\label{sec:optimization}
\begin{figure}
	\includegraphics[width=0.35\textwidth]{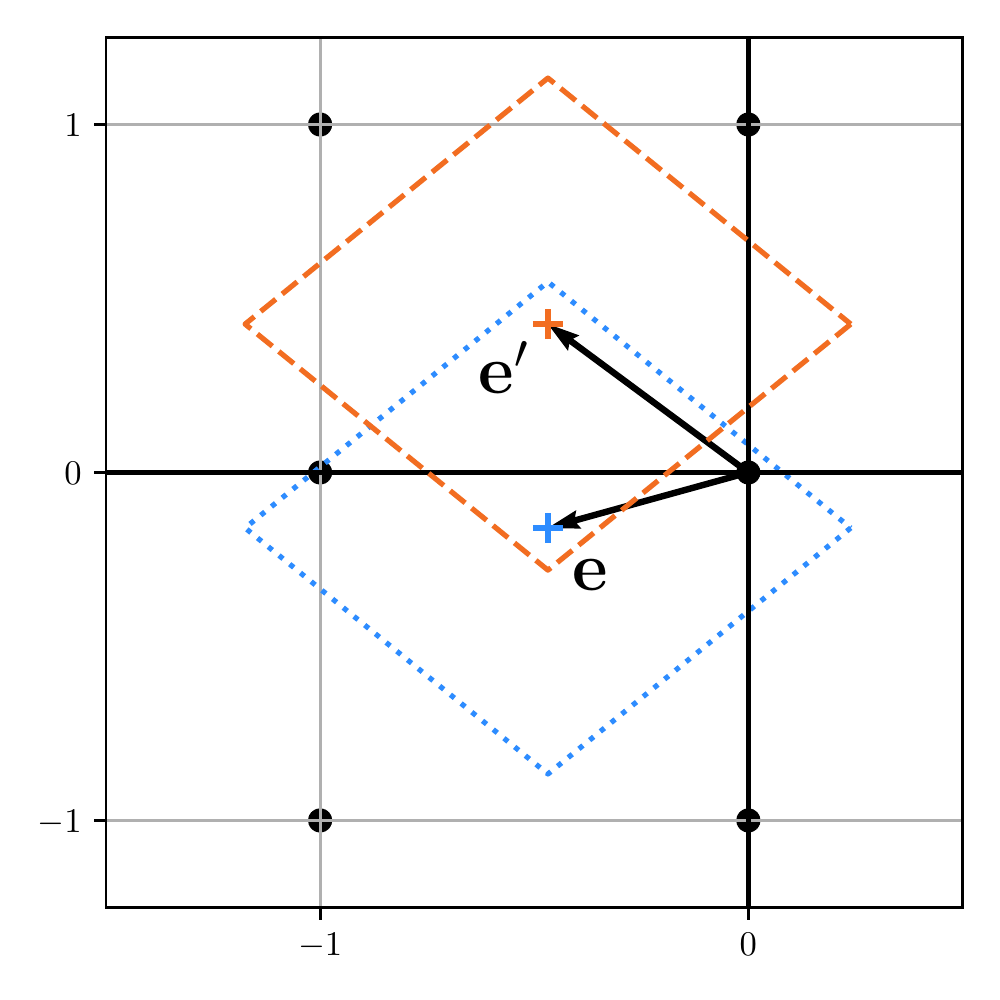}
	\caption{Simple example with toy blocks of size $(1,2)$. $\mathbf{e}$ is the original spatial rounding error but $\mathbf{e'}$ is the spatial rounding error obtained after embedding a message in the DCT coefficient. The blue dotted square and the orange dashed square defines the constraints of the characterization of antecedent for both situations. Note that no compatible integer values are present inside the dashed square.}
	\label{example}
\end{figure}
\begin{figure*}[h]
	\begin{tikzpicture}[node distance=2cm]
	\node (image) [textbox] {JPEG image};
	\node (block1) [textbox, right of=image, yshift=0.25cm, xshift=0.25cm] {Block $(n,m)$};
	\node (block2) [textbox, right of=image, yshift=-0.5cm, xshift=0.25cm] {Block $(n,m)$};
	\node (ilp) [textbox, right of=block1, xshift=0.25cm] {ILP problem};
	\node (gurobi) [textbox, right of=ilp, xshift=0.25cm] {Gurobi solver};
	\node (dot1) [right of=block2, xshift=-0.5cm]{};
	\node (dot2) [right of=dot1, xshift=-0.5cm]{};
	\node (res) [right of=gurobi, xshift=4cm]{
		$\begin{cases}
		\textbf{infeasible} \rightarrow \text{incompatible = stego}   \\
		\textbf{feasible, good solution} \rightarrow \text{continue}  \\
		\textbf{time limit reached, no solution} \rightarrow \text{continue with timing attack}  \\
		\textbf{feasible, wrong solution} \rightarrow \text{ignore, continue}  \\
		\end{cases}$};
	
	\draw [arrow] (image.east) -- (block1.west);
	\draw [arrow] (image.east) -- (block2.west);
	\draw [arrow] (block1.east) -- (ilp.west);
	\draw [arrow] (ilp) -- (gurobi);
	\draw [arrow] (gurobi) -- (res);
	\draw [dotted] (dot1) -- (dot2);
	\end{tikzpicture}
	\caption{Diagram of the model}
	\label{fig:diagramm}
\end{figure*}
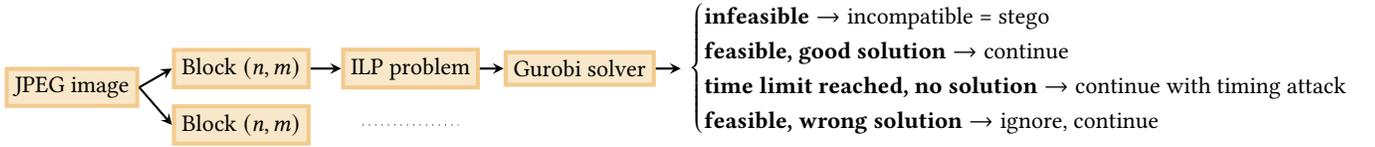

In this section, we formalize the problem of finding an antecedent of a DCT block as a canonical Integer Linear Programming (ILP) problem. Even if such a problem is NP-hard, there exist methods and solvers able to find the solution or prove that the problem is infeasible.

Note also that we do not know the DCT coefficient $\mathbf{d}$ before the compression, but we can use $\mathbf{c}$ instead to have a good approximation of the sign. This approximation can be wrong around 0 and create wrong formulations, however, we can detect the problematic blocks by compressing their antecedent, and if we do not find the same DCT coefficients we know that the problem was incorrect and we ignore this block. This issue only concern a minority of block based on our experiment on $8\times8$ blocks in ALASKA dataset.

With the same notations as before, a potential antecedent $\mathbf{\tilde{x}} = [\mathbf{y}] - \mathbf{k}$ is compatible if $\mathbf{k}$ is solution of the following ILP problem:

\begin{equation}
	\begin{aligned}
	\min_{\mathbf{k}} \quad & 1\\
	\text{s.t.} \quad & \forall\, i, \left\{\begin{aligned}
	-0.5 < \tilde{u}_i \leq 0.5, &\quad \text{ if } c_i \geq 0,\\
	-0.5 \leq \tilde{u}_i < 0.5, &\quad \text{ if } c_i < 0.
	\end{aligned}
	\right.
	\end{aligned}
\end{equation}

Note that the objective function does not matter since we are only interested in feasible solutions to this ILP problem.

Let define $\boldsymbol{\delta}$ a mask vector where $\delta_i = 1$ if $c_i \geq 0$, 0 otherwise. The canonical form can be defined as follows. Details about derivation can be found in appendix \ref{ilp standard form}.

\begin{equation}
\begin{aligned}
\min_{\mathbf{k}} \quad & 1,\\
\text{s.t.} \quad & \lim_{\varepsilon \rightarrow 0}\,  \mathbf{A}\mathbf{k} \leq \mathbf{b}_{\varepsilon},
\end{aligned}
\end{equation}
where $\mathbf{A} = \begin{pmatrix} \mathbf{M} \\ -\mathbf{M}\end{pmatrix} \in \mathbb{R}^{2nm\times nm}$ and $\mathbf{b}_{\varepsilon} = 0.5 + \mathbf{A}\mathbf{e} - \begin{pmatrix} 1-\boldsymbol{\delta} \\ \boldsymbol{\delta}\end{pmatrix} \varepsilon \in \mathbb{R}^{2nm}$ and $\mathbf{k} \in \mathbb{Z}^{nm}$. The limit on $\varepsilon$ is used to transform strict inequalities into inequalities.

\subsection{Example}

To get a visualization of such compatible blocks, we can see the problem as follow. The constraints on $\mathbf{\tilde{u}}$ define a hypercube of size 1 in dimension $nm$. Since the DCT is a rotation, $\mathbf{M}^T\mathbf{\tilde{u}}$ is a rotated hypercube. Therefore, the compatible antecedents are the integer coordinates inside the rotated hypercube of size 1 translated by $\mathbf{e}$. 

We construct a toy example to propose a visual representation of the problem. Therefore, we need to work with blocks of size $(1,2)$. Let $\mathbf{x} = (195,84)$ be the original block. Its DCT coefficients after subtracting 128 are $\mathbf{c} = (16,78)$ and the decompression gives us the spatial block $[\mathbf{y}] = (194,84)$. The approximate spatial rounding error is $\mathbf{e} = (-0.47, -0.16)$. This example is presented in figure \ref{example}. The vector $\mathbf{e}$ is simply the rounding error and the dotted blue cube around it defines the constraints of the characterization. Inside this square (cube in 2D), there are two values of $\mathbf{k}$ which create a compatible antecedent: $(-1,0)$ and $(0,0)$.
Both values of $\mathbf{k}$ are compatible since $\mathbf{x} + (-1,0) = (194,84)$ and $\mathbf{x} + (0,0) = (195,84)$ after compression gives the DCT block $\mathbf{c}$.

We now suppose that we embed a message in the DCT coefficients of the same block. We have a modification vector $\mathbf{m} = (1,-1)$ such that the new DCT block is $\mathbf{c'} = \mathbf{c} + \mathbf{m}$. The decompressed and rounded block is $\mathbf{y'} = (194,86)$ and the new spatial rounding error $\mathbf{e'} = (-0.47, 0.43)$ as shown on the same figure. We can see that for this modified DCT blocks, there is no integer point inside the dashed red square. It means that this DCT block is incompatible. 

The same situation generalizes to other dimensions and in particular $(8,8)$. This example also gives intuitions about other QF. Indeed, lower QF will increase the size of the cube constraint depending on the coefficient inside the quantization table. Thus, the main intuition is that the lower the QF, the more antecedent exist and the less this method can be used for detection. The next section is dedicated to a deeper analysis of incompatible blocks.

\subsection{Solving ILP problem}

Solving ILP problems requires algorithms that belong to the fields of optimization. Fortunately, \texttt{Gurobi}\cite{gurobi} with its python module named \texttt{Gurobipy} can solve such problems. 
The solver is based on a branch-and-bound approach that explores a tree of search. The number of iterations is related to the difficulty of the problem. In particular, for an infeasible problem, the search tree will be larger and deeper in order to prove that no feasible solution exists. In this case, the number of iterations is larger than when the search stops at the first feasible solution. 

The DCT block is decompressed to extract the spatial rounding error used to build the canonical form of the optimization problem as seen in section ~\ref{sec:optimization}. Then, the problem is given to \texttt{Gurobi} which stops its search if any feasible solution to the problem is found or the time limit is reached. In the first case, it means that we found a $\mathbf{k}$ such that $\mathbf{\tilde{u}} = \mathbf{M}(\mathbf{k} - \mathbf{e})$ is a compatible rounding error in the DCT domain that lead to an antecedent in the spatial domain. In this case, we cannot deduce anything from the block. However, if \texttt{Gurobi} detects that the ILP problem is infeasible, it means that no antecedent exists in the spatial domain. We then conclude that the block has been modified in the DCT domain. A diagram of this model is presented in figure \ref{fig:diagramm}. More details on the implementation and the parameters used can be found on the git of the project~\footnote{\url{https://gitlab.cristal.univ-lille.fr/elevecqu/inversing-jpeg-compression-qf100}}.

\section{Analysis of incompatible blocks}

When applying a modification to the DCT coefficients, one can create an incompatible block but it is also possible to obtain another compatible block. The question that arises from this observation is, what is the probability to create an incompatible block? This section proposes two experiments to give answers to the latter question.

In the first experiment, we record the proportion of incompatible blocks for different numbers of modifications and different block sizes. In the second experiment, we record the detection power for all DCT blocks in the images depending on the payload in a sandbox framework. Indeed, images will be compressed in blocks of size $(6,6)$ and embedded using an LSBM algorithm. The reason is that blocks of size $(6,6)$ can be proven incompatible very quickly compared to $(8,8)$ blocks.

Both experiments use blocks cropped from real images present in the ALASKA dataset in pixel format. The JPEG compression is done on each block as explained in section~\ref{sec:jpeg} using the round half away from zero function (i.e. $[0.5] = 1$ and $[-0.5] = -1$).
\begin{figure}
	\includegraphics[width=0.35\textwidth]{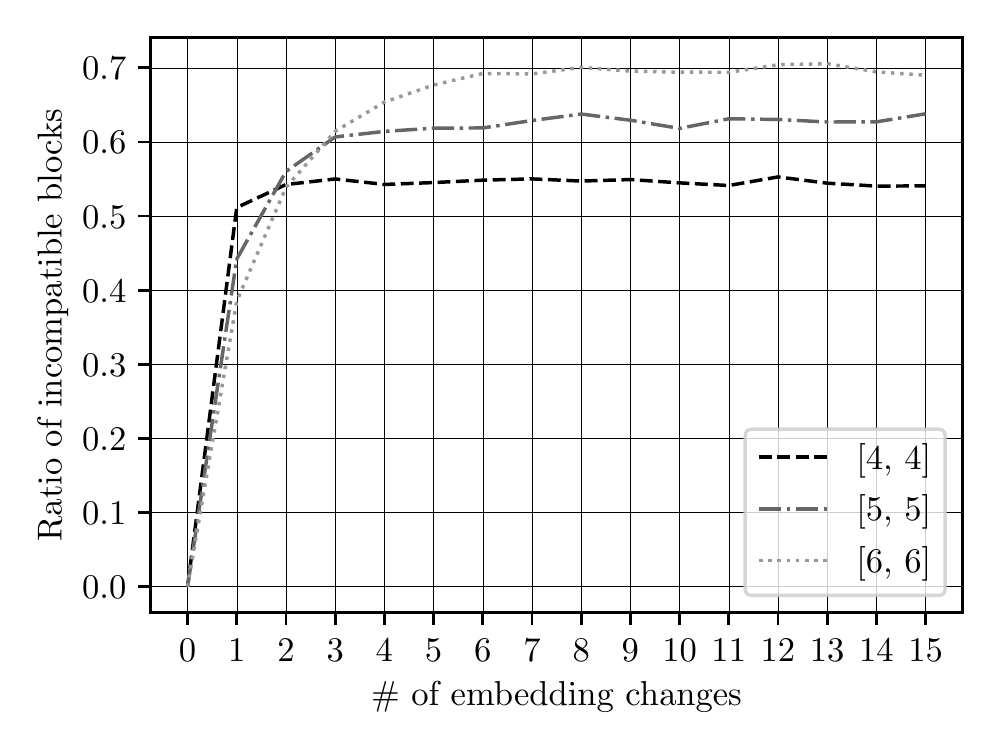}
	\caption{Proportion of incompatible blocks detected depending on the number of modification apply to the DCT coefficients. For block of size $(6,6)$, the empirical probability of building an incompatible block by modifying 4 coefficient is approximately 65\%.}
	\label{fig:expe1}
\end{figure}
\subsection{Number of changes and incompatibility}

In this experiment, we use random blocks of size $(4,4)$, $(5,5)$, and $(6,6)$. For each block, we select $p$ random DCT coefficients and each coefficient is going to be modified by either $+1$ or $-1$. We then build the optimization problem as explained in section~\ref{sec:optimization}. And we use the \texttt{Gurobi} solver through the \texttt{Gurobipy} Python module to try to find a solution to the ILP problem.

Figure \ref{fig:expe1} depicts the empirical probability of creating an incompatible block. The first observation is that the probability is quite high even for a single modification and this probability grows with the number of modifications. The second observation is that the probability depends on the size of the block. For a single modification, the bigger the block, the least probable it is to be incompatible. Last observation, the trend seems to converge to a fixed value after several changes.

We try to give an interpretation of this convergence. For no change at all, the probability equals 0 because there is always an antecedent and every DCT blocks are compatible if no modification has been done. Using the example from figure~\ref{example}, it means that an integer coordinate is inside the cube of constraint. Then it is important to understand that the side of the hypercube is the same but its diagonals gets bigger with the dimension. Thus, applying a small modification in the DCT domain (a single change for example) will translate the block only a little. Hence, there is still a large probability to have the same integer coordinate inside the dashed hypercube. However for more modifications, the translation of the block is bigger, so we converge toward the probability to take a randomly rotated hypercube of side length 1 anywhere in the space and to compute the odd that there is no integer point inside this cube.

We want to draw the reader's attention to the fact that those probabilities are for a single block. Let us assume the worst-case scenario in size $(6,6)$, every modified block of an image has only one modification. The probability to detect such a block is around 0.4 which means that the probability of detecting nothing and thus having a false negative is $(1 - 0.4)^{r}$ with $r$ the number of modified blocks. This value is very small, close to $0.6\%$, for only 10 modified blocks. 

The next experiment is a deeper analysis of this detection power for a given payload.

\subsection{Payload and detection power}

\begin{figure}
	\includegraphics[width=0.35\textwidth]{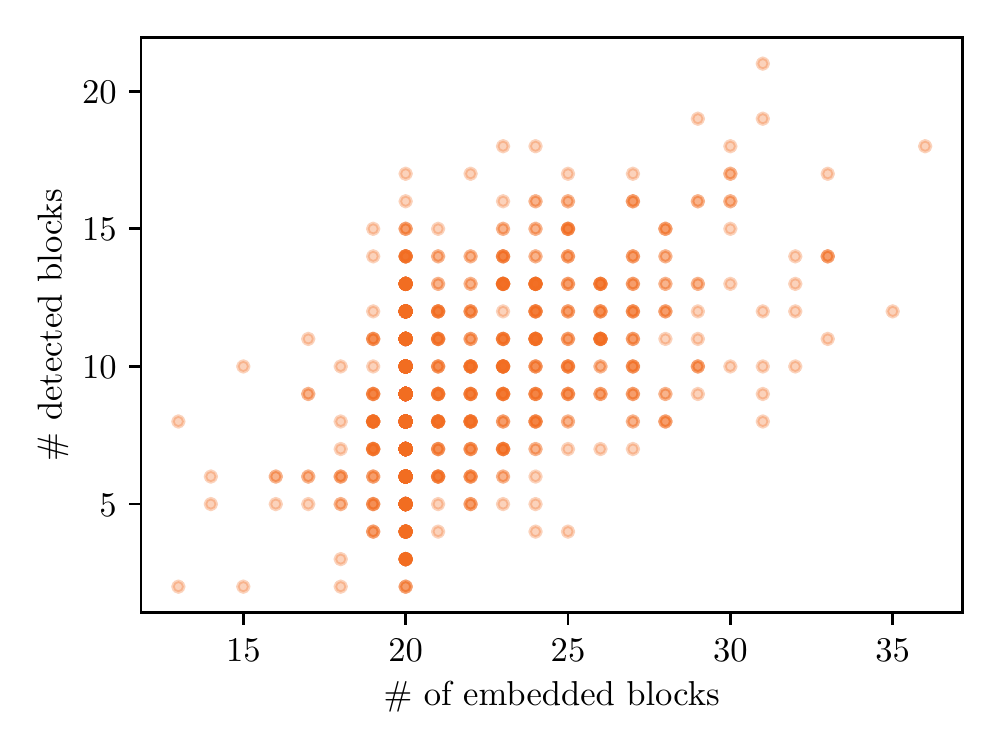}
	\caption{Number of incompatible DCT blocks depending on the number of modified blocks using 1000 images of size $256\times256$ embedded using an LSBM with blocks size of $(6,6)$ and payload 0.01 bpnzac.}
	\label{fig:sandbox}
\end{figure}

For this experiment, we want to evaluate the power detection of such a model. However, proving that a block of size $(8,8)$ is incompatible takes too much time. Therefore, we will be using blocks of size $(6,6)$. Images come from the ALASKA dataset and are of size $256\times256$ but will be cropped to fit a multiple of the block size. We use a modified LSBM algorithm to embed images with a payload of 0.01 bpnzac (bit per non-zero AC coefficient) using optimal coding. Note that contrary to adaptive ternary embedding schemes, LSBM with optimal coding is the strategy which, on average, minimizes the number of embedding changes. Figure \ref{fig:sandbox} presents the number of incompatible blocks depending on the number of embedded blocks for such and embedding rate. This experiment shows that the detection power is very good since even a single incompatible block can prove that the DCT coefficients of the images have been changed. In this experiment, we don't have any errors over 1000 images.

Those sandbox experiments give some understanding of the compatibility however they did not represent the real-world scenario of detecting stego messages embedded in $(8,8)$ DCT blocks. The next section presents a timing attack to address this realistic scenario.

\section{Timing attack on real stego images}

\subsection{Experimental framework}

ILP problems are classified as NP-hard and therefore the complexity to solve them grows very quickly with the number of variables. In our case, this number of variables is the size of blocks. Hence, we can prove that blocks of size $(7,7)$ are incompatible after several minutes but the time required to do the same with blocks of size $(8,8)$ has never been reached. However, when blocks are compatible, the time required to find a feasible solution to the ILP problem is very reasonable. For example, the longest time recorded on our machines is less than 20 minutes to find a solution in the worst case. This difference of time, related to the important number of iterations to solve harder ILP problems, is targeted in this experiment to build a timing attack. 

The ALASKA dataset \cite{cogranne2021alaska} is composed of gray-scale images from different sources, but in our case sources do not matter at all since only JPEG compression at a fixed quality factor is targeted. Images are embedded using J-UNIWARD~\cite{holub2014universal} with a payload of 0.1 bpnzac.

We start by filtering our clipped blocks. For example, if during decompression the block is clipped between 0 and 256, we ignore this block because the rounding error can be outside of our model. For all other blocks, we extract the spatial rounding error and we build an ILP problem as presented in section \ref{sec:optimization}. Using \texttt{Gurobipy} we try to find an antecedent of this block in the spatial domain. However, we stop the solver after a certain time. There are two possibilities from this point, either the solver found a solution in less than the time limit or the solver did not find anything, and the problem is unsolved. 

\subsection{Results}

\begin{figure}
	\includegraphics[width=0.35\textwidth]{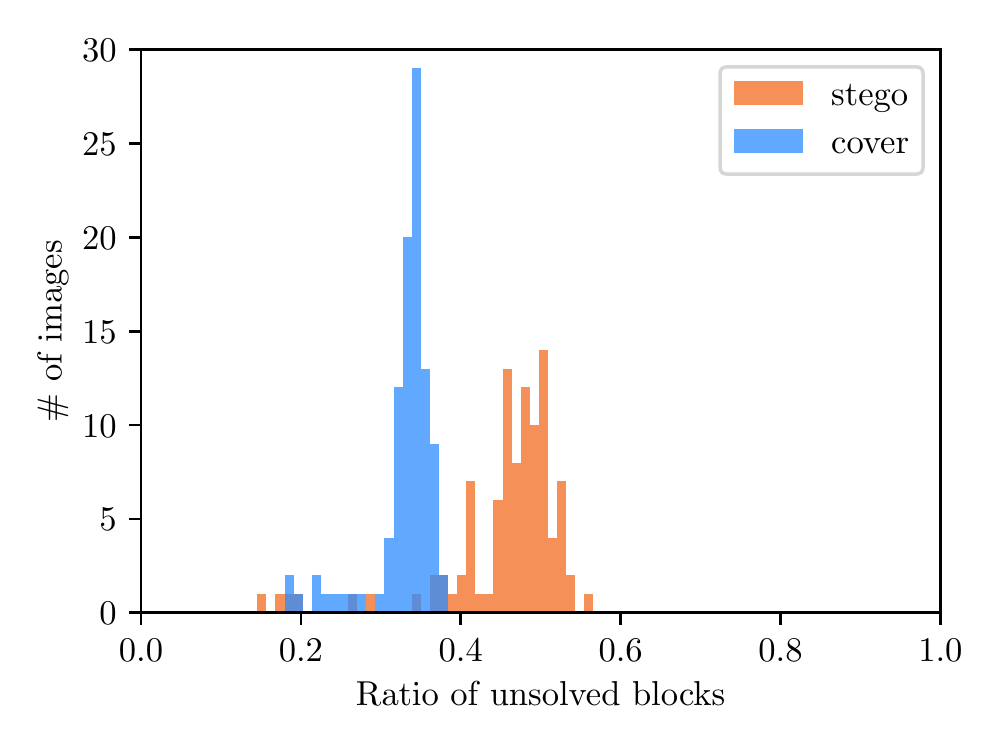}
	\caption{Distribution of the ratio of "unsolved" blocks of size $(8,8)$ per image with a time limit of 100 seconds for each block. "unsolved" blocks are those for which the ILP problem has not been solved.}
	\label{fig:hist}
\end{figure}

\begin{figure}
	\includegraphics[width=0.35\textwidth]{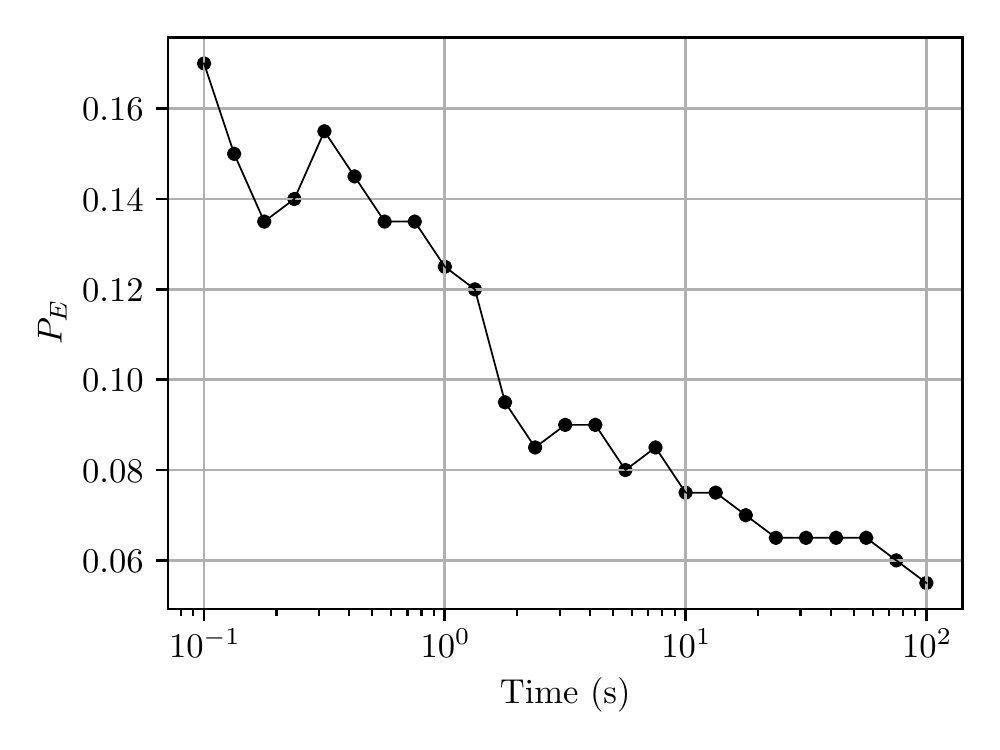}
	\caption{Evolution of the minimal error under equal prior for different time limits for block of size $(8,8)$. The longer the time, the more discriminating the model is.}
	\label{fig:error}
\end{figure}

Figure \ref{fig:hist} shows the distribution of the ratio of unsolved blocks per image for cover and stego images with a time limit of 100 seconds for each block. This ratio of unsolved blocks per image is a very discriminating feature to detect stego images as shown in figure \ref{fig:error} which presents the evolution of the minimal total error under equal priors defined as:
\begin{equation*}
	P_{E} = \min_{P_{FA}} \frac{P_{FA} + P_{MD}(P_{FA})}{2}
\end{equation*}

As the time limit growths, the solver will find feasible solutions for all blocks of cover images. Therefore, the distribution of the ratio of unsolved block per images will collapse to a Dirac in 0. But for stego images, incompatible blocks will remain unsolved and the ratio of unsolved block will converge to the proportion of incompatible blocks in the image. 

We see however that the designed timing attack exbibits an exponential complexity w.r.t. the probability of error, and that it can be rather time consuming: the time had to be multiplied by $100$ in order to divide the probability of error by 2.  

\section{Conclusions and perspectives}

For high quality-factor JPEG image, modifying the DCT coefficients can creates incompatible blocks that do not have any antecedent in the spatial domain. Using a characterization of compatible blocks we derive an ILP problem formulation. With the help of a solver, we can show that some blocks are incompatible for reasonable size up to $(7,7)$. This method is perfectly reliable and benefits from a very good detection power capable of detecting payloads of 0.01 bpnzac. However solving ILP with block size of $(8,8)$ takes too much time but the time can also be used as a feature to discriminate stego images from covers. Using this timing attack, we show that reliable detector can be obtain with sufficient amount of time.

In futures work it could be interesting to have a deeper analysis of this incompatible blocks for lower quality-factor. Indeed, the constraints defined by the characterization should grow larger and thus enabling more solutions to the ILP problem. Another perspective could be to improve the ILP problem and its implementation. There exist symmetry properties that have been ignored in this work and which could be the key to solve infeasible blocks of size $(8,8)$.


\begin{acks}
The work presented in this paper received funding from the European Union’s Horizon 2020 research and innovation program under grant agreement No 101021687 (project “UNCOVER”).

Authors also thanks Tomas Pevny for letting them discover the \texttt{Gurobi} solver and helping them.
\end{acks}

\bibliographystyle{ACM-Reference-Format}
\bibliography{paperIH}


\begin{thebibliography}{10}


\ifx \showCODEN    \undefined \def \showCODEN     #1{\unskip}     \fi
\ifx \showDOI      \undefined \def \showDOI       #1{#1}\fi
\ifx \showISBNx    \undefined \def \showISBNx     #1{\unskip}     \fi
\ifx \showISBNxiii \undefined \def \showISBNxiii  #1{\unskip}     \fi
\ifx \showISSN     \undefined \def \showISSN      #1{\unskip}     \fi
\ifx \showLCCN     \undefined \def \showLCCN      #1{\unskip}     \fi
\ifx \shownote     \undefined \def \shownote      #1{#1}          \fi
\ifx \showarticletitle \undefined \def \showarticletitle #1{#1}   \fi
\ifx \showURL      \undefined \def \showURL       {\relax}        \fi
\providecommand\bibfield[2]{#2}
\providecommand\bibinfo[2]{#2}
\providecommand\natexlab[1]{#1}
\providecommand\showeprint[2][]{arXiv:#2}

\bibitem[\protect\citeauthoryear{Bene\v{s}, Hofer, and B\"{o}hme}{Bene\v{s}
  et~al\mbox{.}}{2022}]%
        {benes2022libjpeg}
\bibfield{author}{\bibinfo{person}{Martin Bene\v{s}}, \bibinfo{person}{Nora
  Hofer}, {and} \bibinfo{person}{Rainer B\"{o}hme}.}
  \bibinfo{year}{2022}\natexlab{}.
\newblock \showarticletitle{Know Your Library: How the Libjpeg Version
  Influences Compression and Decompression Results}. In
  \bibinfo{booktitle}{\emph{Proceedings of the 2022 ACM Workshop on Information
  Hiding and Multimedia Security}} (Santa Barbara, CA, USA)
  \emph{(\bibinfo{series}{IH\&MMSec '22})}. \bibinfo{publisher}{Association for
  Computing Machinery}, \bibinfo{address}{New York, NY, USA},
  \bibinfo{pages}{19–25}.
\newblock
\showISBNx{9781450393553}
\urldef\tempurl%
\url{https://doi.org/10.1145/3531536.3532962}
\showDOI{\tempurl}


\bibitem[\protect\citeauthoryear{Butora and Bas}{Butora and Bas}{2022}]%
        {butora2022compressor}
\bibfield{author}{\bibinfo{person}{Jan Butora} {and} \bibinfo{person}{Patrick
  Bas}.} \bibinfo{year}{2022}\natexlab{}.
\newblock \showarticletitle{{High Quality JPEG Compressor Detection via
  Decompression Error}}. In \bibinfo{booktitle}{\emph{{GRETSI}}}.
  \bibinfo{address}{Nancy, France}.
\newblock
\urldef\tempurl%
\url{https://hal.science/hal-03697777}
\showURL{%
\tempurl}


\bibitem[\protect\citeauthoryear{Butora and Fridrich}{Butora and
  Fridrich}{2019}]%
        {butora2019reverse}
\bibfield{author}{\bibinfo{person}{Jan Butora} {and} \bibinfo{person}{Jessica
  Fridrich}.} \bibinfo{year}{2019}\natexlab{}.
\newblock \showarticletitle{Reverse JPEG compatibility attack}.
\newblock \bibinfo{journal}{\emph{IEEE Transactions on Information Forensics
  and Security}}  \bibinfo{volume}{15} (\bibinfo{year}{2019}),
  \bibinfo{pages}{1444--1454}.
\newblock


\bibitem[\protect\citeauthoryear{Cogranne, Giboulot, and Bas}{Cogranne
  et~al\mbox{.}}{2020}]%
        {cogranne2021alaska}
\bibfield{author}{\bibinfo{person}{R{\'e}mi Cogranne}, \bibinfo{person}{Quentin
  Giboulot}, {and} \bibinfo{person}{Patrick Bas}.}
  \bibinfo{year}{2020}\natexlab{}.
\newblock \showarticletitle{{ALASKA-2 : Challenging Academic Research on
  Steganalysis with Realistic Images}}. In \bibinfo{booktitle}{\emph{{IEEE
  International Workshop on Information Forensics and Security}}}.
  \bibinfo{address}{New York City (Virtual Conference), United States}.
\newblock
\urldef\tempurl%
\url{https://hal-utt.archives-ouvertes.fr/hal-02950094}
\showURL{%
\tempurl}


\bibitem[\protect\citeauthoryear{Fridrich, Goljan, and Du}{Fridrich
  et~al\mbox{.}}{2001}]%
        {fridrich2001steganalysis}
\bibfield{author}{\bibinfo{person}{Jessica Fridrich}, \bibinfo{person}{Miroslav
  Goljan}, {and} \bibinfo{person}{Rui Du}.} \bibinfo{year}{2001}\natexlab{}.
\newblock \showarticletitle{Steganalysis based on JPEG compatibility}. In
  \bibinfo{booktitle}{\emph{Multimedia Systems and Applications IV}},
  Vol.~\bibinfo{volume}{4518}. SPIE, \bibinfo{pages}{275--280}.
\newblock


\bibitem[\protect\citeauthoryear{{Gurobi Optimization, LLC}}{{Gurobi
  Optimization, LLC}}{2023}]%
        {gurobi}
\bibfield{author}{\bibinfo{person}{{Gurobi Optimization, LLC}}.}
  \bibinfo{year}{2023}\natexlab{}.
\newblock \bibinfo{title}{{Gurobi Optimizer Reference Manual}}.
\newblock
\newblock
\urldef\tempurl%
\url{https://www.gurobi.com}
\showURL{%
\tempurl}


\bibitem[\protect\citeauthoryear{Holub, Fridrich, and Denemark}{Holub
  et~al\mbox{.}}{2014}]%
        {holub2014universal}
\bibfield{author}{\bibinfo{person}{Vojt{\v{e}}ch Holub},
  \bibinfo{person}{Jessica Fridrich}, {and} \bibinfo{person}{Tom{\'a}{\v{s}}
  Denemark}.} \bibinfo{year}{2014}\natexlab{}.
\newblock \showarticletitle{Universal distortion function for steganography in
  an arbitrary domain}.
\newblock \bibinfo{journal}{\emph{EURASIP Journal on Information Security}}
  \bibinfo{volume}{2014}, \bibinfo{number}{1} (\bibinfo{year}{2014}),
  \bibinfo{pages}{1--13}.
\newblock


\bibitem[\protect\citeauthoryear{Kocher}{Kocher}{1996}]%
        {kocher1996timing}
\bibfield{author}{\bibinfo{person}{Paul~C Kocher}.}
  \bibinfo{year}{1996}\natexlab{}.
\newblock \showarticletitle{Timing attacks on implementations of
  Diffie-Hellman, RSA, DSS, and other systems}. In
  \bibinfo{booktitle}{\emph{Advances in Cryptology—CRYPTO’96: 16th Annual
  International Cryptology Conference Santa Barbara, California, USA August
  18--22, 1996 Proceedings 16}}. Springer, \bibinfo{pages}{104--113}.
\newblock


\bibitem[\protect\citeauthoryear{Kodovsk{\`y} and Fridrich}{Kodovsk{\`y} and
  Fridrich}{2013}]%
        {kodovsky2013jpeg}
\bibfield{author}{\bibinfo{person}{Jan Kodovsk{\`y}} {and}
  \bibinfo{person}{Jessica Fridrich}.} \bibinfo{year}{2013}\natexlab{}.
\newblock \showarticletitle{JPEG-compatibility steganalysis using
  block-histogram of recompression artifacts}. In
  \bibinfo{booktitle}{\emph{Information Hiding: 14th International Conference,
  IH 2012, Berkeley, CA, USA, May 15-18, 2012, Revised Selected Papers 14}}.
  Springer, \bibinfo{pages}{78--93}.
\newblock


\bibitem[\protect\citeauthoryear{Yousfi, Butora, Fridrich, and Giboulot}{Yousfi
  et~al\mbox{.}}{2019}]%
        {yousfi2019breaking}
\bibfield{author}{\bibinfo{person}{Yassine Yousfi}, \bibinfo{person}{Jan
  Butora}, \bibinfo{person}{Jessica Fridrich}, {and} \bibinfo{person}{Quentin
  Giboulot}.} \bibinfo{year}{2019}\natexlab{}.
\newblock \showarticletitle{Breaking ALASKA: Color separation for steganalysis
  in JPEG domain}. In \bibinfo{booktitle}{\emph{Proceedings of the ACM Workshop
  on Information Hiding and Multimedia Security}}. \bibinfo{pages}{138--149}.
\newblock


\end{thebibliography}

\appendix

\section{Details on the characterization of JPEG antecedents}\label{characterization details}

Even if the characterization has been presented for quality factor 100, it can be extended to every quality factor as it will be shown here. Let $\mathbf{q}$ be the quantization table. The division between two vectors is performed element-wise in the following derivations and the symbol $\cdot$ denotes the element-wise multiplication. The notations need to be extended to take this quantization into account:
\begin{alignat*}{3}
\mathbf{x} & &\in \left[0;255\right]^{nm} \qquad &\text{Original integer block flattened}\\
\mathbf{d} &= \mathbf{M} \mathbf{x} \quad &\in \mathbb{R}^{nm} \qquad &\text{Floating DCT coefficients}\\
\mathbf{c} &= [\mathbf{d} / \mathbf{q}] &\in \mathbb{Z}^{nm} \qquad &\text{Integer DCT coefficients}\\
\mathbf{y} & = \mathbf{M}^T (\mathbf{c} \cdot \mathbf{q}) & \in \mathbb{R}^{nm} \qquad &\text{Floating decompressed block}\\
\mathbf{e} & = [\mathbf{y}] - \mathbf{y}  &\in \mathbb{R}^{nm}\qquad &\text{Rounding error in spatial domain}\\
\mathbf{u} & = [\mathbf{d} / \mathbf{q}] - \mathbf{c}  &\in \mathbb{R}^{nm} \qquad& \text{Rounding error in DCT domain}
\end{alignat*}
Equation \ref{eq1} becomes:
\begin{equation}\label{eq2}
\begin{split}
[\mathbf{y}] - \mathbf{x}& = [\mathbf{y}] - \mathbf{M}^T\mathbf{d}\\
& = [\mathbf{y}] - \mathbf{M}^T((\mathbf{c} - \mathbf{u})\cdot \mathbf{q})\\
& = [\mathbf{y}] - \mathbf{M}^T(\mathbf{c} \cdot \mathbf{q}) + \mathbf{M}^T(\mathbf{u} \cdot \mathbf{q})\\
& = [\mathbf{y}] - \mathbf{y} + \mathbf{M}^T(\mathbf{u} \cdot \mathbf{q}) \\
& = \mathbf{e} + \mathbf{M}^T(\mathbf{u} \cdot \mathbf{q})
\end{split}
\end{equation}
And the characterization becomes:
Let $\mathbf{k} \in \mathbb{Z}^{nm}$. We define $\mathbf{\tilde{x}} = [\mathbf{y}] - \mathbf{k}$ and $\mathbf{\tilde{u}} = (\mathbf{M}(\mathbf{k} - \mathbf{e}))/\mathbf{q}$ the DCT rounding error associated to the compression of $\mathbf{\tilde{x}}$. $\mathbf{\tilde{x}}$ is a compatible antecedent of $\mathbf{c}$ if and only if
\begin{equation}\label{constraints generalized}
\left\{\begin{split}
-0.5 < \tilde{u}_i \leq 0.5, &\quad \text{ if } d_i \geq 0,\\
-0.5 \leq \tilde{u}_i < 0.5, &\quad \text{ if } d_i < 0.
\end{split}\right.
\end{equation}
Proof of the necessary assertion is obtained by construction, we details the calculus for the sufficient assertion. Let $\mathbf{k}$ be a vector of integers such that $\mathbf{\tilde{u}} = (\mathbf{M}(\mathbf{k} - \mathbf{e}))/\mathbf{q}$ respects the constraint in \ref{constraints generalized}. We can define the potential antecedent, $\mathbf{\tilde{x}} = [\mathbf{y}] - \mathbf{k}$. The compression of this antecedent is the following:
\begin{equation*}
\begin{split}
\left[\frac{\mathbf{M} \mathbf{\tilde{x}}}{\mathbf{q}} \right] & = \left[\frac{\mathbf{M}\left([\mathbf{y}] - \mathbf{k}\right)}{ \mathbf{q}}\right] \\
& = \left[\frac{\mathbf{M}\left(\mathbf{y} + \mathbf{e}\right) - \mathbf{M}\left(\mathbf{e} + \mathbf{M}^T\left(\mathbf{\tilde{u}} \cdot \mathbf{q}\right)\right)}{\mathbf{q}}\right] \\
& = \left[\frac{\mathbf{c} \cdot \mathbf{q} - \mathbf{\tilde{u}}\cdot\mathbf{q}}{\mathbf{q}}\right] \\
& = \left[\mathbf{c} - \mathbf{\tilde{u}}\right] \\
& = \mathbf{c}
\end{split}
\end{equation*}
We proved that the compression of an antecedent build thanks to the constraints of the characterization yield the same compressed block thus being compatible and proving the sufficient assertion.

\section{Derivation of the ILP canonical form}\label{ilp standard form}
A potential antecedent $\mathbf{\tilde{x}} = [\mathbf{y}] - \mathbf{k}$ is compatible if $\mathbf{k}$ is solution of the following ILP problem:

\begin{equation}
\begin{aligned}
\min_{\mathbf{k}} \quad & 1\\
\text{s.t.} \quad & \forall\, i, \left\{\begin{aligned}
-0.5 < \tilde{u}_i \leq 0.5, &\quad \text{ if } c_i \geq 0,\\
-0.5 \leq \tilde{u}_i < 0.5, &\quad \text{ if } c_i < 0.
\end{aligned}
\right.
\end{aligned}
\end{equation}

\noindent
Using the problem notations $\mathbf{M}, \mathbf{k}$ and $\mathbf{e}$ the constraints are equivalent to:

\begin{equation}
\forall\, i, \left\{\begin{aligned}
-(\mathbf{M}\mathbf{k})_i &< 0.5 - (\mathbf{M}\mathbf{e})_i  &\text{ if } c_i \geq 0 \\
(\mathbf{M}\mathbf{k})_i &\leq 0.5 + (\mathbf{M}\mathbf{e})_i &\text{ if } c_i \geq 0 \\
-(\mathbf{M}\mathbf{k})_i &\leq 0.5 - (\mathbf{M}\mathbf{e})_i &\text{ if } c_i < 0\\
(\mathbf{M}\mathbf{k})_i & < 0.5 + (\mathbf{M}\mathbf{e})_i &\text{ if } c_i < 0
\end{aligned}
\right.
\end{equation}

\noindent
In order to remove the strict inequalities, one introduces $\varepsilon > 0$ a slack variable going to 0:

\begin{equation}
\forall\, i, \left\{
\begin{aligned}
\lim_{\varepsilon \rightarrow 0} &&-(\mathbf{M}\mathbf{k})_i & \leq 0.5 - (\mathbf{M}\mathbf{e})_i - \varepsilon &\text{ if } c_i \geq 0 \\
&&(\mathbf{M}\mathbf{k})_i &\leq 0.5 + (\mathbf{M}\mathbf{e})_i &\text{ if } c_i \geq 0 \\
&&-(\mathbf{M}\mathbf{k})_i & \leq 0.5 - (\mathbf{M}\mathbf{e})_i &\text{ if } c_i < 0\\
\lim_{\varepsilon \rightarrow 0} &&(\mathbf{M}\mathbf{k})_i &\leq 0.5 + (\mathbf{M}\mathbf{e})_i -\varepsilon &\text{ if } c_i < 0
\end{aligned}
\right.
\end{equation}

\noindent
Finally, let us define $\boldsymbol{\delta}$ a mask vector where $\delta_i = 1$ if $c_i \geq 0$, 0 otherwise. The problem can be defined as:

\begin{equation}
\begin{aligned}
\min_{\mathbf{k}} \quad & 1\\
\text{s.t.} \quad & \lim_{\varepsilon \rightarrow 0}\,  \mathbf{A}\mathbf{k} \leq \mathbf{b}_{\varepsilon}
\end{aligned}
\end{equation}
where $\mathbf{A} = \begin{pmatrix} \mathbf{M} \\ -\mathbf{M}\end{pmatrix}$ and $\mathbf{b}_{\varepsilon} = 0.5 + \mathbf{A}\mathbf{e} - \begin{pmatrix} 1 -\boldsymbol{\delta} \\ \boldsymbol{\delta}\end{pmatrix} \varepsilon$

\end{document}